\documentclass[a4paper,12pt]{article}
\usepackage[utf8]{inputenc}
\pdfoutput=1
\usepackage{jheppub,float}
\usepackage[normalem]{ulem}
\usepackage{xcolor}
\usepackage{amsmath}
\usepackage{framed}
\colorlet{shadecolor}{lightgray}
\usepackage{hyperref}
\usepackage{parskip}

\title{\boldmath Asymptotic Symmetry and Classical holographic dual of $dS_3$ supergravity}

\author{Arindam Bhattacharjee}

\affiliation{Indian Institute of Science Education \& Research Pune,\\
	Homi Bhabha Road, Pashan, Pune 411 008,\\ Maharashtra, India.}
	
\affiliation[]{E-mail: arindam.bhattacharjee@students.iiserpune.ac.in}

\abstract{We consider minimal supergravity on (2+1)dimensional de-Sitter background. We fix the fall-off conditions for gravitini fields in order to fix the asymptotic phase space. Using the Chern-Simons formulation, we then derive the asymptotic symmetry algebra for this theory. The fall-off conditions impose constraints on the phase space which reduces the Chern Simons theory to a WZW model. Further constraints reduce it to a super-Liouville theory at the boundary. This can be treated as a classical dual for the supergravity theory in the bulk.}

\dedicated{Last Updated: \today}

\begin{document} 
\maketitle
\flushbottom

\section{dS/CFT and holographic dual}
The seminal work of Brown and Henneaux \cite{Brown:1986nw} provided us with the precursor of Holographic duality in the case of (2+1)D gravity. Their work suggested that asymptotic symmetry group of locally $AdS_3$ backgrounds give rise to centrally extended Virasoro algebra at the boundary, which is the symmetry algebra for 2D conformal field theories. Following this idea, a whole body of work focused on trying to find a 2D theory at the boundary that exhibited the same symmetry structure \cite{Coussaert:1995zp,Banados:1998pi} . It was shown that upto zero modes, these theories boil down to Liouville theory\cite{Seiberg:1990eb,Teschner:2001rv}. Thus the work \cite{Brown:1986nw} laid the foundation for AdS/CFT.

Taking a flat limit of the $AdS_3$ spacetime, the boundary symmetry algebra of asymptotically flat spaces were found \cite{Barnich:2012aw, Barnich:2006av}. It was the (2+1)D analogue of BMS algebra originally discovered in \cite{Bondi:1962px, Sachs:1962wk}. Subsequent work showed a flat limit of Liouville theory describes the dynamics of these spacetimes \cite{Barnich:2012rz}. The work in both $AdS_3$ and Flat sacetimes were extended to supergravity theories \cite{Coussaert:1993jp,Bautier:1999ds,Barnich:2015sca,Lodato:2016alv,Fuentealba:2017fck,Banerjee:2018hbl,Henneaux:1999ib,Banerjee:2019lrv}. Because of its strong resemblance with the negative cosmological constant case, a BMS/CFT conjecture has recently been put forward in the same spirit\cite{Barnich:2010eb,Bagchi:2012xr}.

\begin{figure}\label{ds-penrose}
\begin{center}
\includegraphics[scale=0.6]{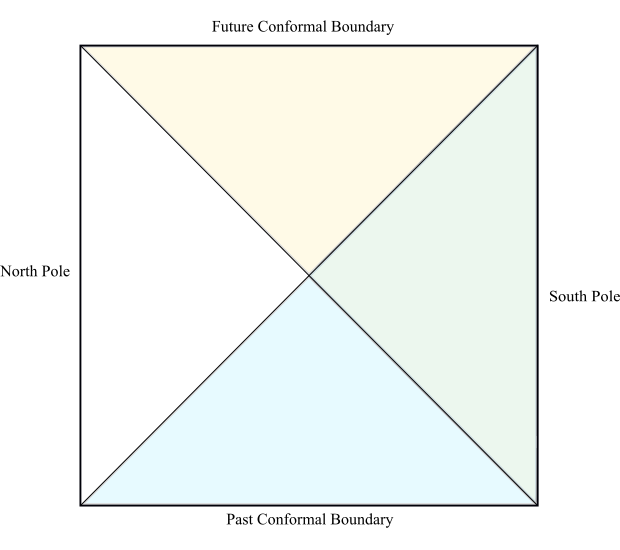}
\end{center}
\caption{Penrose diagram for de-Sitter space. For a static observer at the south pole, the blue region is the causal past and the yellow region is the causal future. The overlapping region is the causal diamond for the observer.}
\end{figure}

Strominger proposed the initial dS/CFT correspondence \cite{Strominger:2001pn} as a natural generalisation of AdS/CFT to positive cosmological constant case. Even in the original work, a big question centred around where the holographic dual should reside. A direct analytic extension of AdS/CFT ideas would suggest that the dual theory lives both in the past($\mathcal{I}^-$) and future boundary($\mathcal{I}^+$) of dS \cite{Cacciatori:2001un}. The reason for the worry is, de-Sitter space has cosmological horizons. Which means, any static observer, for example, is not able to access the whole space-time [Figure \ref{ds-penrose}]. Another related problem in de-Sitter is that with enough matter present, a space-time which is de-Sitter in far past, may collapse in finite time and have no future dS like structure at all. Thus if the holographic dual theory has part of it described on the future boundary, that theory will become meaningless.

~~\\
By an anti-podal matching condition between $\mathcal{I}^+$ and $\mathcal{I}^-$ Strominger was able to show that the holographic description only at one of these is enough. Many other 'screens' for the dual were also proposed. But the situation, like many other problem, greatly simplifies in (2+1)D as was shown in \cite{Compere:2014cna}. They showed, that due to the property of radial gauge choice, it is possible to bring the holographic dual theory in the static patch of the observer itself. Although in principle this enables us to write a consistent boundary theory at any radial slice, its more natural to impose the boundary conditions to a slice close to the horizon.\\
~~\\
In this work, we generalise the above construction to minimal supergravity in $dS_3$ background while working in Eddington-Finkelstein co-ordinate. We extend the idea of 'asymptotically dS spacetime' to supergravity by giving consistent fall-offs for gravitinis. This extends the asymptotic symmetry group, which now lies in the radial slice discussed above. Then using hamiltonian reduction, we write the classical holographic dual for this supergravity theory which turns out to be a euclidean super-Liouville theory at the boundary. \\

This paper is organised as follows: In section 2 we briefly review the minimal super-algebra that has de-Sitter symmetry group as the bosonic subgroup. We write this algebra in a suitable basis and discuss it's supertraces which is essential to write the CS theory corresponding to supergravity. In section 3, we find out the fall-off of metric and gravitini fields and calculate the asymptotic symmetry algebra. Finally in section 4, we use the constaints of the theory to write a boundary dual for the theory.
 
\section{de-Sitter supergravity in (2+1)D}
Absence of local degrees of freedom makes gravity topological in (2+1)D. It was shown \cite{Witten:1988hc} that gravity action in (2+1)D can be written in terms of Chern-Simons theory. The gauge group of the CS theory is the local isometry group of the underlying manifold. Since, gravity is topological, the underlying manifold is locally just flat, AdS or dS depending on the value of the cosmological constant.\\
When dealing with a supergravity theory, the gauge group is chosen to be the supergroup containing the isometry group. In order to write the action, it's also necessary that the (super)group under consideration has non-degenerate invariant bilinears.

\subsection{de-Sitter algebra and it's supersymmetrization}
In a generic dimension $d$, the symmetry algebra of de-Sitter space is given by $\mathfrak{so}(d,1)$. In order to have a supersymmetric extension of this algebra, we must find a super-algebra whose bosonic subalgebra is of the form $\mathfrak{so}(d,1) \oplus \mathfrak{g}$. In his work Nahm \cite{Nahm:1977tg} classified all such algebras. In particular the super-algebra $OSp(N|2,\mathbb{C})$ has the bosonic subgroup $\mathfrak{so}(N)\times \mathfrak{sl}_2(\mathbb{C})$. Hence $\mathcal{N} = 1$ de-Sitter supergravity is possible in (2+1)D.  \\
~~\\
The algebra $OSp(1|2,\mathbb{C})$ in a particular basis looks like\cite{Koehler:1990vh}:
\begin{align*}
&[J^a, J^b] = \epsilon^{abc} J_c; \qquad
[P^a, J^b] = \epsilon^{abc} P_c; \qquad
[P^a, P^b] = -\lambda \epsilon^{abc} J_c \\
&[J^a, U_{\alpha}] = -(\sigma^a)^{\beta}_{\alpha} U_{\beta} \qquad
[J^a, V_{\alpha}] = -(\sigma^a)^{\beta}_{\alpha} V_{\beta} \qquad
[P^a, U_{\alpha}] = \sqrt{\lambda}(\sigma^a)^{\beta}_{\alpha} V_{\beta} \qquad
[P^a, V_{\alpha}] = -\sqrt{\lambda}(\sigma^a)^{\beta}_{\alpha} U_{\beta}\\
&\{U_{\alpha}, U_{\beta} \} = 2 (\sigma^a \epsilon_{\alpha\beta})  P_a \qquad
\{V_{\alpha}, V_{\beta} \} = -2 (\sigma^a \epsilon_{\alpha\beta}) P_a \qquad
\{U_{\alpha}, V_{\beta} \} = 2 \sqrt{\lambda} (\sigma^a \epsilon_{\alpha\beta})  J_a \qquad
\end{align*}
It has 6 bosonic generators $\{P_a, J_a\}$ with $a = \{0,1,2\}$ and 4 fermionic generators $\{U_{\alpha}, V_{\alpha}\}$ where $\alpha = \{-,+\}$. The $\lambda$ parameter appearing in the algebra is inversely proportional to radius of de-Sitter space and $\lambda \rightarrow 0$ limit gives $\mathcal{N} = (1,1)$ super-Poincar\'{e} algebra as discussed, for example, in \cite{Banerjee:2019lrv}.\\
~~\\
But for our purposes this basis poses a problem. So, we'll go to a different basis of this algebra which will also make the $\mathfrak{sl}_2(\mathbb{C})$ structure of the algebra apparent. To do so, we define the following new combinations:
\begin{align*}
J^{\pm}_{a} = \frac{1}{2} (J_a \pm \frac{i}{\sqrt{\lambda}}P_a) \qquad \qquad Q_{\alpha} =\frac{1}{2\lambda^{1/4}}(U_{\alpha} - i V_{\alpha}) \qquad \bar{Q}_{\alpha} =\frac{1}{2\lambda^{1/4}} (U_{\alpha} + i V_{\alpha})
\end{align*}
In this new basis the algebra looks like:
\begin{align}\label{ds-susy}
&[J^+_a, J^+_b] = \epsilon_{ab}^c J^+_c \qquad[J^+_a, Q_{\alpha}] = -(\sigma^a)^{\beta}_{\alpha} Q_{\beta}
 \qquad
\{Q_{\alpha},Q_{\beta}  \} = -i (\sigma^a\epsilon_{\alpha\beta})J^+_a
 \qquad \nonumber\\
&[J^-_a, J^-_b] = \epsilon_{ab}^c J^-_c \qquad
[J^-_a, \bar{Q}_{\alpha}] = -(\sigma^a)^{\beta}_{\alpha} \bar{Q}^-_{\beta} \qquad
\{\bar{Q}_{\alpha},\bar{Q}_{\beta}  \} = i (\sigma^a\epsilon_{\alpha\beta})J^-_a
 \qquad \nonumber\\
&[J^+_a, J^-_b] = 0; \qquad
[J^+_a, \bar{Q}_{\alpha}] = [J^-_a, Q_{\alpha}] = 0 \qquad
\{Q_{\alpha},\bar{Q}_{\beta}  \} = 0
\end{align}
~~\\
In our conventions $\varepsilon_{012} = 1$ and the tangent space metric is given by:
$\eta_{ab}=
\begin{pmatrix}
0 & 1 & 0\\
1 & 0 & 0\\
0 & 0 & 1\\
\end{pmatrix}$.
The $\sigma^a$ matrices satisfy $[\sigma^a, \sigma^b] = \epsilon^{abc}\sigma_c$.\\
~~~\\
In the basis (\ref{ds-susy}), we can define invariant bilinears of the algebra. In general, this will be a one-parameter family but the parameter will be fixed by demanding that the bosonic part of the CS action reduces to Einstein gravity action.
The non-zero supertraces are given by:
\begin{eqnarray}
\langle J^+_a, J^+_b\rangle =  \eta_{ab} \quad
\langle Q_{\alpha}, Q_{\beta}\rangle = i C_{\alpha\beta} \quad
\langle J^-_a, J^-_b\rangle = \eta_{\alpha\beta} \quad
\langle \bar{Q}_{\alpha}, \bar{Q}_{\beta}\rangle = -i C_{\alpha\beta}
\end{eqnarray}

It must be mentioned that the algebra above actually admits one more invariant bilinear which corresponds to an independent set of supertraces. But this choice does not correspond to a non-degenerate metric as $\lambda \rightarrow 0$.
\subsection{Writing the supergravity action}
In this above basis we can expand the CS gauge field as:
\begin{align}
A &= (\omega^a + i\sqrt{\lambda} e^a) J^+_a  + \lambda^{1/4}(\psi^{\alpha} + i \chi^{\alpha})Q_{\alpha}  \nonumber \\
\bar{A} &=  (\omega^a - i\sqrt{\lambda} e^a) J^-_a + \lambda^{1/4}(\psi^{\alpha} - i \chi^{\alpha})\bar{Q}_{\alpha}
\end{align}
Then the Einstein action can be written as a sum of CS actions of $A$ and $\bar{A}$.
\begin{align}\label{EH action}
S_{EH} = -i S_{\kappa}[A] +i S_{\kappa} [\bar{A}]
\end{align}
where $S_{\kappa}[A]$ is the Chern-Simons action given by:
\begin{align*}
S_{\kappa}[A] = \frac{\kappa}{2} \int_{\mathcal{M}} \left\langle A \wedge dA + \frac{2}{3} A \wedge A \wedge A \right\rangle
\end{align*}
The Chern-Simons level $\kappa$ is given by $\kappa = \frac{1}{8G}$. Like gravity in (2+1)D, the Chern-Simons theory is also topological. It's equations of motion are just $F = 0$, where $F$ is the field strength given by $F = dA + A \wedge A$. This means that the field $A$ is locally pure gauge and the solution can be written in the form $A = G^{-1}d G$, where $G$ is a $SL(2,\mathbb{C})$ element.\\
~~\\
In terms of the component fields the equations of motion becomes:
\begin{align}
de^c &+ \varepsilon_{ab}^{\,\, \, \, c} e^{a} \omega^b + (\sigma^c \epsilon)_{\alpha\beta} (\psi^{\alpha} \psi^{\beta} - \chi^{\alpha} \chi^{\beta}) = 0 \nonumber \\
d\omega^c &+ \frac{1}{2} \varepsilon_{ab}^{\,\, \, \, c} (\omega^{a} \omega^b - \lambda e^{a} e^{b}) + 2 \sqrt{\lambda} \psi^{\alpha} \chi^{\beta} (\sigma^c \epsilon)_{\alpha\beta} = 0 \nonumber \\
d\psi^{\beta} &- \omega^a \psi^{\alpha} (\sigma^a)^{\beta}_{\alpha} - \sqrt{\lambda} e^{a} \chi^{\alpha}(\sigma^a)^{\beta}_{\alpha} = 0 \nonumber \\
d\chi^{\beta} &- \omega^a \chi^{\alpha} (\sigma^a)^{\beta}_{\alpha} + \sqrt{\lambda} e^{a} \psi^{\alpha}(\sigma^a)^{\beta}_{\alpha} = 0
\end{align}
this we present here just for completeness. For $\lambda \rightarrow 0$ limit these equations get decoupled and a generic solution is possible \cite{Banerjee:2019lrv}.

\section{Asymptotic symmetry of dS Supergravity}
Since gravity is topological in (2+1)D, the boundary conditions play a pivotal role in determining it's behaviour. These conditions give the action a proper variation and also introduces global degrees of freedom. So our main goal is to define consistent boundary conditions for dS supergravity and then analyze the asymptotic symmetry algebra that accompanies it.\\

\subsection{Fall off for field $A$ and $\bar{A}$}
We want to translate the boundary condition from the language of the metric and gravitini field to the gauge field $A$ and $\bar{A}$. Let us briefly review (following \cite{Compere:2014cna}) the phase space of asymptotically dS spacetimes and then extend it to supergravity.\\
In Eddington-Finkelstein co-ordinates the fall-off for metric is given by:
\begin{align}
ds^2 = \left(\frac{r^2}{l^2}+8G\mathcal{M}(u,\phi)\right) du^2 -2du\,dr +8G\mathcal{J}(u,\phi) du\, d\phi + r^2 d\phi^2
\end{align}
Putting this in Einstein's equations tell us that the functions $\mathcal{M}$ and $\mathcal{J}$ can be expanded as:
\begin{align}
\mathcal{M} = \mathcal{L}_+ (t_+) + \mathcal{L}_- (t_-) \qquad \mathcal{J} =il (\mathcal{L}_+ (t_+) - \mathcal{L}_- (t_-))
\end{align}
where the new co-ordinates are defined as $t^{\pm} = u \mp il\phi$.\\
Now we may calculate the vielbeins and spin connections from the above metric. Our tangent space metric is given under (\ref{ds-susy}) and in that basis the gauge fields take the form:
\begin{align}
a_{bos} &= (\frac{i}{l}J^+_1 + \frac{i8G}{l}\mathcal{L}_{-}(t_-) J^+_0)dt_{-} \nonumber \\
\bar{a}_{bos} &= (-\frac{i}{l} J^-_1 - \frac{i}{l} 8G \mathcal{L}_+(t_+) J^-_0) dt_{+}
\end{align}
where we have already taken out the radial dependence from both $A$ and $\bar{A}$ using the relation $a_{bos} = k^{-1}(r)(d+A_{bos})k(r)$. Since we'll work in a constant r slice, this reduced form $a_{bos}$ and $\bar{a}_{bos}$ are going to be our dynamical inputs. Also notice that our form is a bit different from that of \cite{Compere:2014cna} because of the difference in conventions.\\
~~\\
Having found out the fall off for the bosonic part, we want to extend it to the full supergravity field \cite{Banados:1998pi}. The bosonic part contains the information of the physical metrics that should be included in the theory. In order to find the appropriate boundary conditions for the fermionic fields as well, we take $a_{bos}$ and $\bar{a}_{bos}$ and act the whole dS supergroup on it. This generates new terms in the boundary and the full field looks like:
\begin{align}\label{fall-off}
&a = (\frac{i}{l}J^+_1 + \frac{i8G}{l}\mathcal{L}_{-}(t_-) J^+_0+ \frac{8G}{l} r^{-}(t_-)Q_{-})dt_{-}\nonumber \\
&\bar{a} = (-\frac{i}{l} J^-_1 - \frac{i}{l} 8G \mathcal{L}_+(t_+) J^-_0 + \frac{8G}{l} \bar{r}^-(t_+) \bar{Q}_-) dt_+
\end{align}\\
The functions appearing in the expression \ref{fall-off} are assumed to have a smooth behaviour at the boundary. Of these, the functions $r^{-}$ and $\bar{r}^-$ are Grassmann valued. The rest are scalar.\\
The boundary condition on these fields can be divided into two categories. Namely,\\
(1)\label{constraint1} $A_{t_+} = 0$ and $\bar{A}_{t_-} = 0$\\
(2a) $ a_{J_2} = a_{Q_+} = 0; \quad a_{J_1} = \frac{i}{l}$ \qquad
(2b) $ \bar{a}_{J_2} = \bar{a}_{Q_+} = 0; \quad \bar{a}_{J_1} = -\frac{i}{l}$\\
~~\\
We'll shortly see that these boundary conditions will reduce the boundary dual theory first to a WZW action and then finally to a Liouville type theory.\\
But before that we'll analyse the asymptotic symmetry algebra that leaves these conditions invariant.

\subsection{Asymptotic symmetry algebra}
In order to obtain the asymptotic algebra we find the variation of the fields that keeps the above structure intact. The generic variation of the field is given by:
\begin{align}\label{gauge-eqn}
\delta a_{\phi} = d\, \lambda + [a_{\phi},\lambda]
\end{align}
where $\lambda$ is the gauge transformation parameter. We expand it in terms of our generators as:
\begin{align*}
\lambda = \xi^a J^+_a + \theta^{\alpha} Q_{\alpha}
\end{align*}
and then use our asymptotic field expression (\ref{fall-off}) to find the variations in (\ref{gauge-eqn}). Matching the coefficients on both sides, we see that not all parameters are independent and are be related by:
\begin{align}
\xi^{2} &= -(\xi^1)' \nonumber\\
\xi^{0} &= - (\xi^1)'' + 8G \mathcal{L}_{-} \xi^1 + 4G r^{-}\theta_{+}\nonumber\\
\theta^{-} &= \sqrt{2} (\theta^+)' - i\,8G r^- \xi^1
\end{align}
Thus the only independent parameters are $\xi^1$ and $\theta^{+}$. Next, we will write the variations of the fields appearing in the expression (\ref{fall-off}). In terms of independent parameters these are given by:
\begin{align}
\delta \mathcal{L}_- &= -\frac{1}{8G} (\xi^1)''' + 2 \mathcal{L}_- (\xi^1)' + \mathcal{L}'_- (\xi^1) + \frac{3}{2} r^- (\theta^+)'+ \frac{1}{2} (r^-)' \theta^+\\
\delta r^- &= i\frac{\sqrt{2}}{8G}(\theta^+)'' + \frac{3}{2} r^- (\xi^1)' + (r^-)' \xi^1 - i\frac{1}{\sqrt{2}} \mathcal{L}_- \theta^+
\end{align}
\\
~~\\
Now we can construct the conserved charge associated with the variations that preserve the boundary condition. For CS theory this charge is given by:
\begin{equation}
\delta Q = \frac{\kappa}{2\pi}\int \langle \lambda,\delta a_{\phi} \rangle
\end{equation}
where the above integration is performed over a constant r slice. With our boundary condition the expression reduces to
\begin{equation}
Q = \int (\xi^1 \mathcal{L}_- + \theta^+ r^-)
\end{equation}
where the equivalence between the CS level and Newton's constant was used. With this charge the asymptotic symmetry algebra can be written as:
\begin{align*}
\{ Q[\lambda_1], Q[\lambda_2] \}_{PB} = \delta_{\lambda_2} Q[\lambda_1]
\end{align*}
With this and the variations of fields given above, the algebra becomes:
\begin{align}
\{\mathcal{L}(\theta), \mathcal{L}(\theta') \} &= \frac{1}{8G} \delta'''(\theta-\theta') - (\mathcal{L}(\theta)+ \mathcal{L}(\theta')) \delta'(\theta-\theta')\nonumber \\
\{\mathcal{L}(\theta), r^-(\theta') \} &= -( r^-(\theta) +\frac{1}{2}  r^-(\theta')) \delta'(\theta-\theta')\\
\{r^-(\theta), r^-(\theta') \} &= -\frac{i}{\sqrt{2}} \mathcal{L}(\theta)\delta(\theta-\theta') + \frac{\sqrt{2}i}{8G} \delta''(\theta-\theta')\nonumber
\end{align}

~~\\
In the barred sector, the story runs in parallel. The parameter relations are given by:
\begin{align*}
\bar{\xi}^2 = -(\bar{\xi}^1)'\\
\xi^0 = -(\bar{\xi}^1)'' + 8G\mathcal{L}_+ \xi^1 + 4G \bar{r}^{-}\bar{\theta}^+\\
\bar{\theta}^- = \sqrt{2} (\bar{\theta}^+)' + i\,8G \bar{r}^- \bar{\xi}^1
\end{align*}
and with these, the variation of fields become:
\begin{align}
\delta \mathcal{L}_+ &= -\frac{1}{8G} (\bar{\xi}^1)''' + 2 \mathcal{L}_+ (\bar{\xi}^1)' + \mathcal{L}'_+ (\bar{\xi}^1) + \frac{3}{2} \bar{r}^- (\bar{\theta}^+)'+ \frac{1}{2} (\bar{r}^-)' \bar{\theta}^+\\
\delta \bar{r}^- &= -i\frac{\sqrt{2}}{8G}(\bar{\theta}^+)'' + \frac{3}{2} \bar{r}^- (\bar{\xi}^1)' + (\bar{r}^-)' \bar{\xi}^1 + i\frac{1}{\sqrt{2}} \mathcal{L}_+ \theta^+
\end{align}
These then produce the asymptotic algebra of the barred sector. The classical poisson brackets are given by:
\begin{eqnarray}
\{\bar{\mathcal{L}}(\theta), \bar{\mathcal{L}}(\theta') \} = -\frac{1}{8G} \delta'''(\theta-\theta') - (\bar{\mathcal{L}}(\theta)+ \bar{\mathcal{L}}(\theta')) \delta'(\theta-\theta')\\
\{\bar{\mathcal{L}}(\theta), \bar{r}^-(\theta') \} = -( \bar{r}^-(\theta) +\frac{1}{2}  \bar{r}^-(\theta')) \delta'(\theta-\theta')\\
\{\bar{r}^-(\theta), \bar{r}^-(\theta') \} = +\frac{i}{\sqrt{2}} \mathcal{L}(\theta)\delta(\theta-\theta') - \frac{\sqrt{2}i}{8G} \delta''(\theta-\theta')
\end{eqnarray}
\\
\section{Dual theory at the boundary}
Now that we have analysed the asymptotic algebra, we want to write a classical dual of the supergravity theory we are considering at the boundary of the spacetime. Now, since we are working on the static patch of the dS spacetime, the boundary in this case will be an $r=const.$ hypersurface as $r\rightarrow \infty$ in Eddington-Finkelstein co-ordinates.\\
Using a hamiltonian reduction of the CS theory at the bulk, we will first write a dual WZW type theory at the boundary. This procedure uses the constraint (1) mentioned earlier. Since this constraint essentially is same for pure gravity and supergravity, this procedure goes through in exactly the same way. But we present it here anyway.
\subsection{Dual WZW model at boundary}
To see how the Chern-Simons action gives rise to super-WZW theory at the boundary we employ the techniques discussed in \cite{Compere:2014cna}. In fact since the first constraint takes the same form in pure gravity and supergravity, the reduction is exactly the same. Generically, the relations between these two theories were discussed in \cite{Elitzur:1989nr}.\\
To illustrate this let us expand the CS theory explicitly in terms of our co-ordinates.
\begin{align}
S_{\kappa}[A] = \frac{\kappa}{2} \int_{\mathcal{M}} dr\,du\,d\phi  \langle A_r (\dot{A_{\phi}} - A_{u}') - A_u (\partial_r A_{\phi} - \partial_{\phi} A_{r}) + A_{\phi} (\partial_r A_u - \dot{A_r}) + 2 A_u [A_{\phi}, A_r]\rangle
\end{align}
Where the dots are derivatives w.r.t $u$ and dashes are derivatives w.r.t $\phi$ variables.
Now integrating by parts and taking the $\phi$ boundary terms to be zero because of periodicity the above action upto trivial boundary terms reduces to
\begin{align}
S_{\kappa}[A] = \frac{\kappa}{2} \int_{\mathcal{M}} dr\,du\,d\phi \langle A_r \dot{A_{\phi}} - A_{\phi} \dot{A_r} + 2 A_u F_{\phi r}\rangle
\end{align}
The field strength above is defined as $F_{\phi r} = \partial_{\phi}A_r - \partial_r A_{\phi} + [A_{\phi}, A_r]$. Now we can compute the variation of the above action. We see that apart from the terms proportional to equations of motion ($F = 0$) we also get boundary terms proportional to fields. These come from the commutator in the last term. 
\begin{align}
S_{\kappa}[A] = \int_{\mathcal{M}} \delta(A) (E.O.M) +{\kappa} \int_{\partial \mathcal{M}} du\,d\phi \langle A_u \delta A_{\phi} \rangle
\end{align}
At this point we can use our first set of boundary conditions $A_{t_+} = 0$ at $r \rightarrow \infty$. This implies that at the boundary $A_{u} = A_{\phi}$. Thus the correct action must be supplemented with an additional boundary term to have a well defined variational principle. The action takes the form:
\begin{align*}
S_{\kappa}[A] = \frac{\kappa}{2} \int_{\mathcal{M}} \left\langle A \wedge dA + \frac{2}{3} A \wedge A \wedge A \right\rangle + \frac{\kappa}{2} \int_{\partial \mathcal{M}} du\,d\phi \langle A^2_{\phi} \rangle
\end{align*}
Now similarly the barred Chern-Simons theory will also be supplemented with a boundary term. The final action takes the form:
\begin{align}
S_{IEH} = S_{EH} + i\frac{\kappa}{2}  \int_{\partial \mathcal{M}} du\,d\phi \langle A^2_{\phi} + \bar{A}^2_{\phi} \rangle
\end{align} 
where $S_{IEH}$ is our notation for improved Einstein-Hilbert action, whose variation gives us the correct equations of motion with our boundary conditions. $S_{EH}$ is given by [\ref{EH action}].\\

Since the field strength of the CS action is trivial the fields $A$ and $\bar{A}$ can be expanded as pure gauge
\begin{align}
A = H^{-1}dH \qquad \qquad \bar{A} = \bar{H}^{-1}d\bar{H}
\end{align}
where $H,\bar{H} \in SL(2,\mathbb{C}) $. It's important to understand that the boundary mentioned in the above action isn't the conformal boundary of de-Sitter but rather the boundary of the static patch where we want to see the dual boundary theory. The authors of \cite{Compere:2014cna} have shown that the idea of asymptotic symmetry can be extended to anywhere inside the bulk in the case of (2+1)D pure gravity. It seems natural to put the boundary at the boundary of causal diamond of the static observer.\\
We impose the radial gauge on $H,\bar{H}$ as we've done for the field $A$ earlier. This decomposes the field into $H = h(u,\phi)\,k(r)$ where all the r dependence is now encapsulated in $k$. We do a similar gauge choice for the barred sector $\bar{H} = \bar{h}\,\bar{k}$.\\
With this decomposition the unbarred part of the action (alongwith it's corresponding boundary term) reduces to,
\begin{align}
S_{\kappa}[A] = \frac{\kappa}{2}\left[\int_{\partial \mathcal{M}} \langle h^{-1} \partial_{\phi}h\, h^{-1}\partial_{+}h \rangle + \frac{1}{3}\int_{\mathcal{M}} \langle (H^{-1}dH)^3\rangle \right]
\end{align}
where $\partial_{+}= \partial_{u} - \frac{i}{l}\partial_{\phi}$. This is a chiral WZW model. The barred part reduces similarly to a WZW model of opposite chirality (with $\partial_{-}= \partial_{u} + \frac{i}{l}\partial_{\phi}$)

\begin{align}
S_{\kappa}[\bar{A}] = \frac{\kappa}{2}\left[\int_{\partial \mathcal{M}} \langle \bar{h}^{-1} \partial_{\phi}\bar{h}\, \bar{h}^{-1}\partial_{-}\bar{h} \rangle + \frac{1}{3}\int_{\mathcal{M}} \langle (\bar{H}^{-1}d\bar{H})^3\rangle \right]
\end{align}

To finally reduce these into a non-chiral WZW model, we define
\begin{align}
G = H^{-1}\bar{H} \qquad and \qquad g = h^{-1}\bar{h}
\end{align}
With this identification, the combination of the barred and unbarred CS theory boils down to a non-chiral WZW model:
\begin{align}\label{WZW action}
S_{\kappa}[A] = \frac{\kappa}{2}\left[\int_{\partial \mathcal{M}} \langle g^{-1} \partial_{-}g\, g^{-1}\partial_{+}g \rangle + \frac{1}{3}\int_{\mathcal{M}} \langle (G^{-1}dG)^3\rangle \right]
\end{align}
In the next section, we will impose the rest of the constraints on this action and reduce this boundary dual to a Liouville theory.

\subsection{Super-Liouville action at the boundary}
Now that we've used the first set of boundary conditions to reduce the dual holographic theory at the boundary into a WZW model [\ref{WZW action}], we want to implement the second set of constraints.\\
To do so, we closely follow the construction of \cite{Forgacs:1989ac} and expand do gauss decomposition of the elements of the supergroup close to identity. A generic element then can be written as:
\begin{align*}
g = G^{+}\,G^0\,G^{-}
\end{align*}
where,
\begin{align}\label{gauss-decom}
G^{+}&= \exp(x\Gamma_{1} + \psi_{+}Q^{+})\nonumber\\
G^{-}&= \exp(y\Gamma_{0} + \psi_{-}Q^{-})\\
G^{0}&= \exp(\phi\Gamma_2)\nonumber
\end{align}
In general the fields $x,y,\psi_{\pm},\phi$ will be functions of $(r,u,\theta)$ but due to our gauge choice at the boundary the $r$ dependence is absent. With this form of of the element, we can now calculate the currents of the WZW theory. The theory has 2 independent currents given by:
\begin{align}
J = g^{-1}dg \qquad \bar{J} = - dg\, g^{-1}
\end{align}
Now since in the previous section we've seen that $g = h^{-1}\bar{h}$ we can substitute this in the above expression and get:
\begin{align}
J = \bar{a} - g^{-1}\,a\,g \qquad \bar{J} = a - g\,\bar{a}\,g^{-1}
\end{align}
Here we use the first set of constraints once again and we see that 
\begin{align*}
J_{+} &= \bar{a}_{+} - g^{-1}\,a_{+}\,g\\
     &= \bar{a}_{+}   
\end{align*}
as $a_{+} = 0$. Similarly for the barred sector $\bar{J}_{-} = a_-$. Thus the constraints on the fields can be directly imposed on the components of currents of the theory as well. Using (\ref{gauss-decom}) we now expand the current $J$ in terms of the unbarred basis. We get\\

\begin{align}
J_+ =& [e^{\phi}(\partial_+x) - \frac{i}{\sqrt{2}} e^{\phi} (\partial_+ \psi_+) \psi_+] J^+_1\nonumber\\
&+[-e^{\phi}(\partial_+x)y + \frac{i}{\sqrt{2}}y e^{\phi} (\partial_+ \psi_+) \psi_+ - \frac{i}{2}e^{\phi/2}(\partial_+ \psi_+) \psi_- + \partial_+ \phi ] J^+_2\nonumber\\
&+[(\partial_+ y) - (\partial_+ \phi)y - \frac{i}{2}e^{\phi/2}y(\partial_+ \psi_+) \psi_- + \frac{i}{\sqrt{2}} (\partial_+ \psi_-) \psi_-] J^+_0\nonumber\\
&+[-\frac{1}{\sqrt{2}} e^{\phi} (\partial_+ x) \psi_- + e^{\phi/2} (\partial_+ \psi_+)+ \frac{i}{2} e^{\phi}(\partial_+ \psi_+) \psi_+ \psi_- ]Q_+\nonumber\\
&+[\frac{1}{2} e^{\phi} (\partial_+ x) \psi_- + \frac{1}{\sqrt{2}} e^{\phi/2}y (\partial_+ \psi_+)+ \frac{i}{2\sqrt{2}} e^{\phi}y (\partial_+ \psi_+) \psi_+ \psi_- + \frac{1}{2} (\partial_+ \phi) \psi_- + (\partial_+ \psi_-)]Q_-
\end{align}
In this expression, using the constraints above we find following relations:
\begin{align}\label{confield1}
&e^{\phi} (\partial_+ x)-\frac{i}{\sqrt{2}}e^{\phi} (\partial_+ \psi_+) \psi_+ = \frac{i}{l}\nonumber\\
&y = -\frac{l}{2} e^{\phi/2} (\partial_+ \psi_+) \psi_- - il(\partial_+ \phi)\nonumber\\
&\psi_- = -i \sqrt{2} l e^{\phi/2} (\partial_+ \psi_+)\nonumber\\
\end{align}
Similarly, we expand the $\bar{J}$ current in barred basis and get the expression:
\begin{align}
\bar{J}_- =& -[e^{\phi}(\partial_-y)] J^-_1\nonumber\\
&+[-e^{\phi}x(\partial_- y)  - \frac{i}{2}e^{\phi/2}(\partial_- \psi_-) \psi_+ + \partial_- \phi ] J^-_2\nonumber\\
&-[(\partial_- x) - (\partial_- \phi)x - \frac{i}{2}e^{\phi/2}x(\partial_- \psi_-) \psi_+] J^-_0\nonumber\\
&-[\frac{1}{\sqrt{2}} e^{\phi} (\partial_- y) \psi_+ + e^{\phi/2} (\partial_- \psi_-)]\bar{Q}_+\nonumber\\
&-[(\partial_- \psi_+) - \frac{1}{\sqrt{2}} e^{\phi/2}x (\partial_- \psi_-) - \frac{1}{2} e^{\phi}x (\partial_- y) \psi_+ + \frac{1}{2} (\partial_+ \phi) \psi_-  ]\bar{Q}_-
\end{align}

Then imposing the constraints here gives:
\begin{align}\label{confield2}
&e^{\phi}(\partial_- y) = \frac{i}{l}\nonumber\\
&x = -il (\partial_- \phi) - \frac{l}{2} e^{\phi/2} (\partial_-\psi_-) \psi_+ \nonumber\\
&\psi_+ = i \sqrt{2} l e^{\phi/2} (\partial_-\psi_-)
\end{align}

Thus we see that in the theory the fields $x$ and $y$ can be completely substituted by the fields $\phi,\, \psi_+,\, \psi_-$ and the final action of the constrained theory will be written in terms of these.\\
~~\\
Now after substituting these [\ref{confield1},\ref{confield2}] into the original action improved with the boundary term, we get the Super-Liouville action at the boundary:
\begin{align}
S_E = \kappa \int_{\partial bulk} &[2(\partial_+ \phi)(\partial_- \phi) + \frac{1}{l^2} (e^{2\phi} + \frac{1}{\sqrt{2}} e^{\phi} \psi_+ \psi_-)\nonumber\\
&+ \psi_+ \partial_- \psi_+ + \psi_- \partial_+ \psi_-] dx_+ \, dx_-
\end{align}
This can be treated as the classical boundary dual of the bulk supergravity theory.

\section*{Conclusion and Remarks}
In this work, we discussed the boundary behaviour of supergravity theory in (2+1)D with $\Lambda > 0$. The asymptotic super-algebra was found which contains the bulk isometry group as a subgroup. Finally a classical dual was found which, like the case for pure gravity, is Liouville type theory with an intermediate WZW action. This theory is only dual to the bulk gravity upto holonomy terms and zero modes. It's importance lies in the fact that explicit realisations of the asymptotic symmetry generators can be constructed in terms of the fields of this boundary theory. One may also study the "spectral flow" of algebras as was done for instance in \cite{Henneaux:1999ib}.\\
Another interesting problem to study is the quantum corrections to the asymptotic algebras and the dual theory. For instance the work \cite{Merbis:2019wgk} found non-trivial corrections to the central charges of the boundary algebra. It would be interesting to see these corrections for the boundary theory presented here as well. The problem of considering non trivial holonomies of the bulk gravity theory is also interesting.

\section*{Acknowledgements}
We would like to thank Nabamita Banerjee, Sunil Mukhi, Sachin Jain and Arun Thalapillil for extremely useful discussions regarding this work. We would also like to thank IISER Bhopal for their hospitality. Finally we would like to acknowledge the work of doctors and health workers worldwide for steering us through the COVID-19 pandemic.

\end{document}